\documentclass[lettersize,journal]{IEEEtran}
\usepackage{amsmath,amsfonts,amssymb}
\usepackage{algorithmic}
\usepackage{algorithm}
\usepackage{array}
\usepackage{textcomp}
\usepackage{stfloats}
\usepackage{url}
\usepackage{caption}
\usepackage{subcaption}
\usepackage[hidelinks,colorlinks=true,linkcolor=blue,citecolor=blue,urlcolor=black]{hyperref}
\usepackage{graphicx}
\usepackage{verbatim}
\usepackage{graphicx}
\usepackage{mwe}
\usepackage{enumerate,color}
\usepackage{cite}
\usepackage{acronym}
\usepackage{amssymb} 
\input{AcronymList}
\usepackage{mathtools, nccmath}
\usepackage{subcaption}
\DeclarePairedDelimiter{\nint}\lfloor\rceil

\begin{document}

\title{An Efficient Low-Complexity RSMA Scheme for Multi-User Decode-and-Forward Relay Systems}

\author{
    \IEEEauthorblockN{Ahmet Sacid S\"{u}mer}
    \href{https://orcid.org/0000-0001-8866-8520}{\includegraphics[scale=0.01]{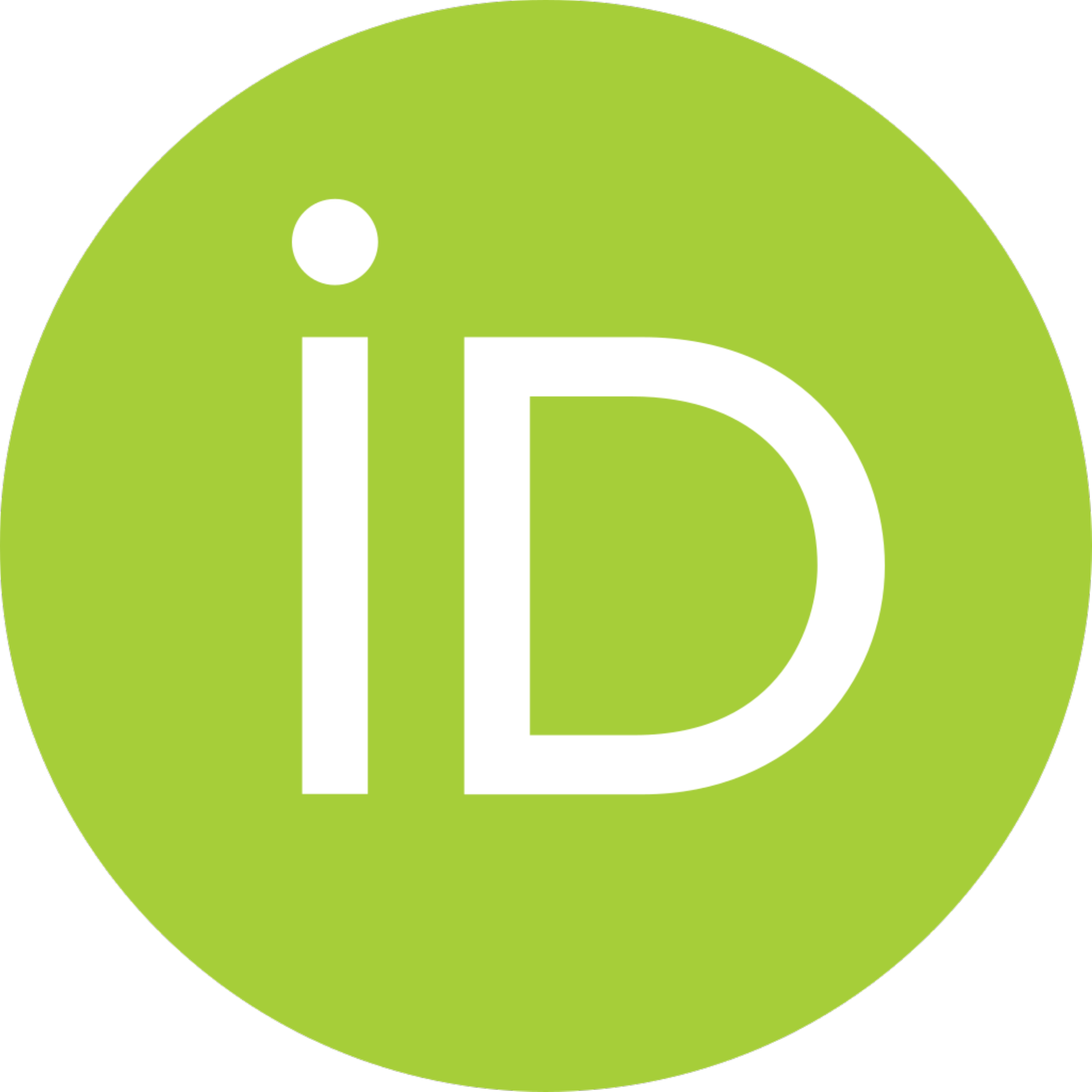}},
    \IEEEauthorblockN{Mehmet Mert \c{S}ahin}
    \href{https://orcid.org/0000-0001-5401-2427}{\includegraphics[scale=0.01]{1024px-ORCID_iD.pdf}},
	\IEEEauthorblockN{and}
	\IEEEauthorblockN{H\"{u}seyin Arslan}
    \href{https://orcid.org/0000-0001-9474-7372}{\includegraphics[scale=0.01]{1024px-ORCID_iD.pdf}},
	\IEEEmembership{Fellow, IEEE}



\thanks{A. S. S\"{u}mer and H. Arslan are with the Department of Electrical and Electronics Engineering, Istanbul Medipol University, Istanbul, 34810, Turkey (e-mail: ahmet.sumer1@medipol.edu.tr; huseyinarslan@medipol.edu.tr). \\
M. M. \c{S}ahin is with Samsung Research America, Samsung Electronics Co., Ltd., Plano, Texas, U.S.A. (e-mail: m.sahin@samsung.com). \\
}
}

\markboth{Journal of \LaTeX\ Class Files,~Vol.~14, No.~8, August~2021}%
{Shell \MakeLowercase{\textit{et al.}}: A Sample Article Using IEEEtran.cls for IEEE Journals}

\maketitle

\begin{abstract}
Rate-Splitting Multiple Access (RSMA) is a promising strategy for ensuring robust transmission in multi-antenna wireless systems. 
In this paper, we investigate the performance of RSMA in a downlink Decode-and-Forward (DF) relay scenario under two phases with imperfect Channel State Information (CSI) at the transmitter and the relay.
In particular, in the first phase, the Base Station (BS) initially transmits to both BS Users (BUs) and the relay. In the second phase, the relay decodes and forwards the received signals to Relay Users (RUs) outside the BS coverage area.
Furthermore, we investigate a scenario where the relay broadcasts a common stream intended for the RUs in the second phase. Due to the broadcast nature of the transmission, this stream is inadvertently received by both the RUs and the BUs. Concurrently, the BS utilizes Spatial Division Multiple Access (SDMA) to transmit private streams to the BUs, resulting in BUs experiencing residual interference from the common stream transmitted from relay.
Incorporating this residual common stream interference into our model results in a significant enhancement of the overall sum-rate achieved at the BUs.
We derive a tractable lower bound on the ergodic sum-rates, 
enables us to develop closed-form solutions for power allocation that maximize the overall sum-rate in both phases.
Extensive simulations validate that our proposed power allocation algorithm, in conjunction with a low-complexity precoder, significantly improves the sum-rate performance of DF relay RSMA networks compared to the SDMA-based benchmark designs under imperfect
CSI at the transmitter and relay.

\end{abstract}
\begin{IEEEkeywords}
Decode-and-forward relays, ergodic sum-rate, linear precoding, MU-MIMO, rate splitting multiple access (RSMA)
\end{IEEEkeywords}
\section{Introduction}
\IEEEPARstart{T}{he} relentless pursuit of ubiquitous, high-speed connectivity in 6G and beyond demands innovative solutions capable of extending the coverage of wireless networks far beyond the capabilities of current technologies. While \ac{MU-MIMO} has been a cornerstone of 4G and 5G, enabling significant gains through beamforming and interference management, its reliance on accurate \ac{CSIT} poses a formidable challenge. In practical scenarios, where channel estimation errors and feedback delays lead to imperfect \ac{CSIT}, \ac{MU-MIMO} struggles to deliver its full potential \cite{jiang2021road}. Conventional approaches to address imperfect \ac{CSIT}, such as dynamic mode switching or sophisticated prediction techniques, often fall short in maintaining seamless multi-user connectivity and scalability, particularly in the face of escalating network demands \cite{saad2019vision}.

\ac{DF} relay-assisted communication emerges as a powerful tool to overcome the coverage and capacity limitations of traditional wireless networks. By strategically deploying intermediate relay nodes, it harnesses the benefits of spatial diversity to combat the detrimental effects of fading and shadowing, significantly extending the coverage of the network \cite{laneman2004cooperative}. 
Going beyond simple signal forwarding, \ac{CDRT}, built on the principles of two-way relaying, further enhances the spectral efficiency of these systems \cite{thai2011coordinated}. In contrast to conventional one-way relaying, where the relay operates solely as a forwarding agent, \ac{CDRT} empowers the relay to actively participate in the communication process. This active participation enables the relay to simultaneously transmit and receive information, effectively utilizing the channel resources and boosting the overall spectral efficiency. 

In this evolving landscape, \ac{RSMA} emerges as a compelling alternative, poised to address the inherent vulnerabilities of relay-assisted communication \cite{clerckx2023primer}. Traditional relaying schemes, while extending coverage, often grapple with the complexities of multi-hop channel estimation and feedback, leading to imperfect \ac{CSIT} and increased susceptibility to \ac{MUI}. \ac{RSMA}, with its strategic splitting of messages into common and private parts coupled with \ac{SIC} at the receivers, offers a robust solution to these challenges \cite{abidrabbu2024ir}. By proactively accounting for imperfect \ac{CSIT} and \ac{MUI} in its design, \ac{RSMA} demonstrates remarkable resilience in scenarios where traditional relaying schemes falter. This inherent robustness makes \ac{RSMA} a natural fit for the demanding environment of relay-assisted communication, where the cascade of channel impairments across multiple hops can significantly degrade performance \cite{hoymann2012relaying}.

\subsection{Related Works}
Several studies have explored the application of \ac{RS} in relay-assisted networks, particularly under various channel conditions and system configurations. 
In \cite{zhang2019cooperative}, the authors explored a cooperative relaying scheme in a two-user relay system, where the user experiencing a higher channel gain acts as a relay to facilitate the transmission of the common message to the user with a weaker channel. Other studies have investigated relaying strategies in multi-antenna broadcast channels, where centrally located users assist the base station in relaying common messages to cell-edge users with perfect \ac{CSIT} \cite{mao2020max}, whereas \cite{zhang2021cooperative} focused on more realistic scenarios with imperfect \ac{CSIT}.
Additionally, \ac{RS} in full-duplex relays has been studied in \cite{khisa2022full}, showcasing its ability to enhance spectral efficiency by allowing \acp{RU} to operate in full-duplex mode.

From a resource allocation perspective, Pang et al. \cite{pang2022resource} aimed to achieve max-min fairness by maximizing the minimum achievable rate in a two-user cooperative \ac{RS} system. Their approach involved jointly optimizing the precoding vectors at both the \ac{BS} and relay, along with time allocation coefficients and common stream rates, under system resource constraints. In \cite{xiao2024intelligent}, a dynamic \ac{RS} scheme was proposed where the near user is adaptively switched on or off during the two transmission phases, improving far user throughput while maintaining near user performance. Similarly, in \cite{papazafeiropoulos2018rate}, \ac{RS} was applied to address rate saturation issues in massive MIMO full-duplex systems, with particular focus on mitigating pilot contamination and self-interference, demonstrating significant improvements in practical settings.

A key limitation of the aforementioned studies lies in their reliance on complex optimization-based methods, which require high computational effort, reducing their feasibility for real-time implementation. To overcome this challenge, more practical solutions have been proposed, such as simplifying the design with low-complexity precoders like \ac{ZF}, coupled with efficient power allocation strategies for \ac{RS}. In particular, several works have examined the use of simpler precoders in both underloaded \cite{clerckx2019rate,hao2015rate,dai2016rate,dai2017multiuser,underloaded,dey2023rsma} and overloaded \cite{overloaded} \ac{RSMA} systems.  In underloaded \ac{RSMA} networks,  the sum-rate performance of 1-layer \ac{RSMA} in a two-user \ac{MISO} broadcast channel using \ac{ZF} precoding for private streams has been analyzed under perfect \ac{CSIT} \cite{clerckx2019rate} and imperfect \ac{CSIT} due to quantization error \cite{hao2015rate}, resulting in a power allocation method that balances power between common and private streams. In massive MIMO systems with imperfect \ac{CSIT}, power allocation between common and private streams has been explored in \cite{dai2016rate}, where a closed-form power allocation algorithm was derived based on the asymptotic assumption of large-scale antenna arrays. For \ac{mmWave} systems, \cite{dai2017multiuser} proposed a hybrid precoding design using second-order channel statistics for \ac{RSMA}, while \cite{underloaded} addressed mobility challenges by employing low-complexity \ac{RSMA} precoding, showing that \ac{RSMA} enhances system resilience against \ac{CSIT} imperfections caused by mobility. Further advancing the application of \ac{RSMA}, \cite{dey2023rsma} examined its potential in massive MIMO systems with mobile \ac{URLLC} users. The authors derived closed-form expressions and developed low-complexity power allocation algorithms to maximize energy efficiency, highlighting \ac{RSMA}’s ability to perform well in challenging channel conditions. 
In overloaded \ac{RSMA} networks, \cite{overloaded} proposed a low-complexity precoder selection and power allocation scheme to achieve max-min fairness under both perfect and imperfect \ac{CSIT}, further illustrating \ac{RSMA}’s scalability in resource-constrained environments.

\subsection{Motivation and Contributions}
Despite the extensive research on relay systems on \ac{RSMA}, prior studies have not addressed 2-hop transmissions under imperfect channel conditions, particularly the challenge of deriving closed-form solutions for power allocation coefficients across two phases with an arbitrary number of antennas and users. To address this gap, this paper investigates the implementation of \ac{RSMA} in multi-user \ac{DF} relay systems, focusing on scenarios with imperfect \ac{CSIT}. The contributions of this paper are as follows:
\begin{itemize}
    \item We propose a multi-antenna \ac{RSMA} system where the \ac{BS} serves single-antenna \acp{BU} and \acp{RU} with relay assistance. 
    In particular, RSMA's ability to broadcast the common stream is utilized, enabling \acp{BU} to receive it during both the first phase and relay transmission, thus enabling cooperative relaying in the second phase. Simultaneously, the BS transmits private streams to \acp{BU}, forming a secondary RSMA scheme to boost spectral efficiency.
    We derive an expression for the residual common stream interference based on the undecodable portion of the common stream rate at the \acp{BU} during the second phase, enabling partial \ac{MUI} decoding at the \acp{BU}.
    
    \item 
    We formulate the total sum-rate optimization problem for \ac{RSMA} in \ac{DF} relay-assisted networks with imperfect \ac{CSIT}, considering multiple \acp{BU} and \acp{RU}. Two relay transmission strategies are evaluated: one where the relay re-encodes common and private streams, and another where it reallocates the common stream rate. These are compared to a scenario without relay sum-rate limitations on \acp{RU}, assuming a Rician fading channel for the relay.
    
    \item A tractable lower bound for the approximated \ac{ESR} of a 1-layer \ac{DF} relay-assisted \ac{RSMA} system is derived by taking into account various \ac{CSIT} imperfections and allowing for an arbitrary number of transmit antennas at the BS and relay, as well as an arbitrary number of users, in both transmission phases. Low-complexity precoders for the private streams are employed to maintain practicality, effectively constraining the design space of formulated problem. Leveraging the derived lower bounds, we obtain a closed-form solution for the power allocation coefficients that maximize the sum-rate performance across both transmission phases. To our knowledge, this is the first paper to propose closed-form power allocation algorithms for a two-phase \ac{DF} relay-assisted \ac{RSMA} system.
    
    \item Our results demonstrate that \ac{RSMA}, coupled with our proposed closed-form power allocation algorithm, effectively mitigates the detrimental effects of imperfect \ac{CSIT} in 2-hop relay-assisted transmission, outperforming \ac{SDMA}. The proposed low-complexity design achieves significant performance gains across various system settings and parameters, showcasing its benefits in both small-scale and massive MIMO regimes, accommodating any number of \acp{BU} and \acp{RU}.
\end{itemize}

The subsequent sections of this paper are organized as follows. Section \ref{section:System-model} provides a description of the system and channel models for the \ac{DF} relay scheme utilizing \ac{RSMA}. In section \ref{sec:proposed}, we introduce our proposed approach, including the transmission model and expressions for the total sum-rate. Section \ref{sec:lowerbound} is dedicated to the derivation of lower bounds for the \acp{ESR} in both phases. Building upon these derived lower bounds, Section \ref{section:Closed Form} presents the derivation of closed-form expressions.
Section \ref{section:Simulation Results} includes a numerical evaluation that demonstrates the achievable sum rate of the proposed strategy.
Finally, Section \ref{section:Conclusion} concludes the paper and outlines potential avenues for future work.\footnote{\textit{Notation:} Vectors and matrices are denoted by boldface lower and upper case symbols, e.g., $\mathbf{a}$ and $\mathbf{A}$. 
The operations $|\cdot|$ and $\|\cdot\|$ denote the absolute value of a scalar, and the $\ell_2$-norm of a vector, respectively.
$(\cdot)^T$, $(\cdot)^H$, $\text{Tr}(\cdot)$, and $\mathbb{E}(\cdot)$ represent the transpose, Hermitian transpose, trace and expectation operators, respectively. $\mathcal{CN}(0, \sigma^2)$ denotes the \ac{CSCG} distribution with zero mean and variance $\sigma^2$. $\mathbf{I}$ denotes the identity matrix.
$\lfloor \cdot \rceil$ denotes the rounding operation.
$\text{Gamma}(D, \theta)$ represents the Gamma distribution with the probability density function $f(x) = \frac{1}{\Gamma(D)\theta^D} x^{D-1}e^{-\frac{x}{\theta}}$  For any complex $x$ with a positive real part, $\Gamma(x) = \int_{0}^{\infty} t^{x-1}e^{-t}dt$ is the gamma function and $\Gamma'(x)$ is its derivative with respect to $x$.}

\begin{figure*}
        \centering
        \begin{subfigure}[b]{0.475\textwidth}
            \centering
\includegraphics[width=\textwidth]{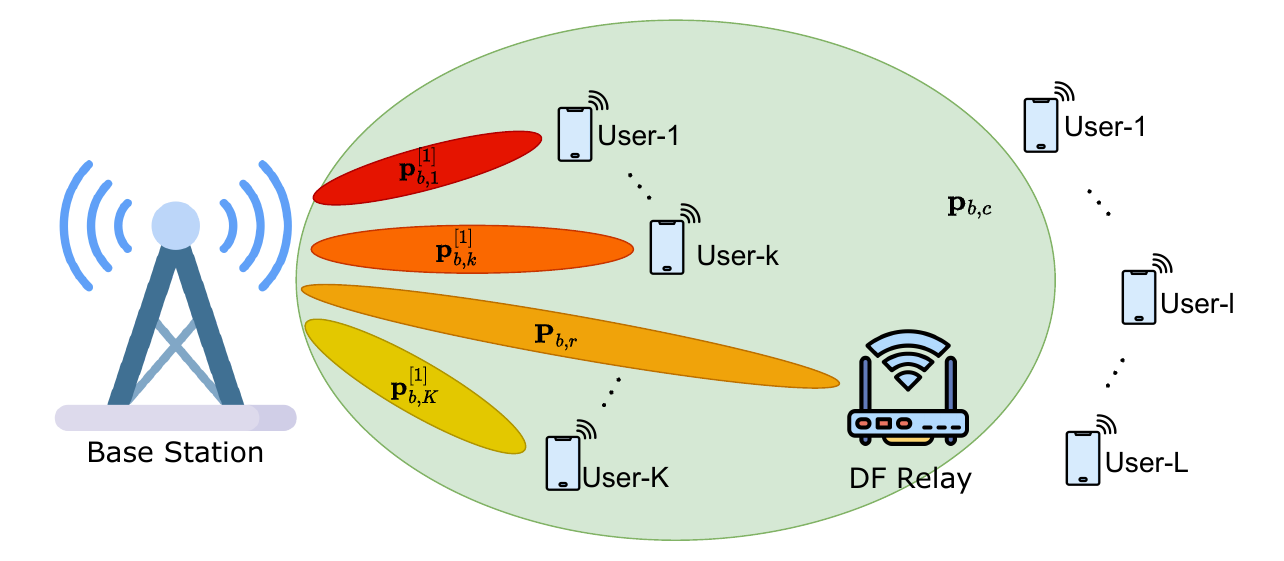}
            \caption[First phase.]%
            {{\small First phase.}} 
            \label{subfig:firstphase}
        \end{subfigure}
        \hfill
        \begin{subfigure}[b]{0.475\textwidth}   
            \centering \includegraphics[width=\textwidth]{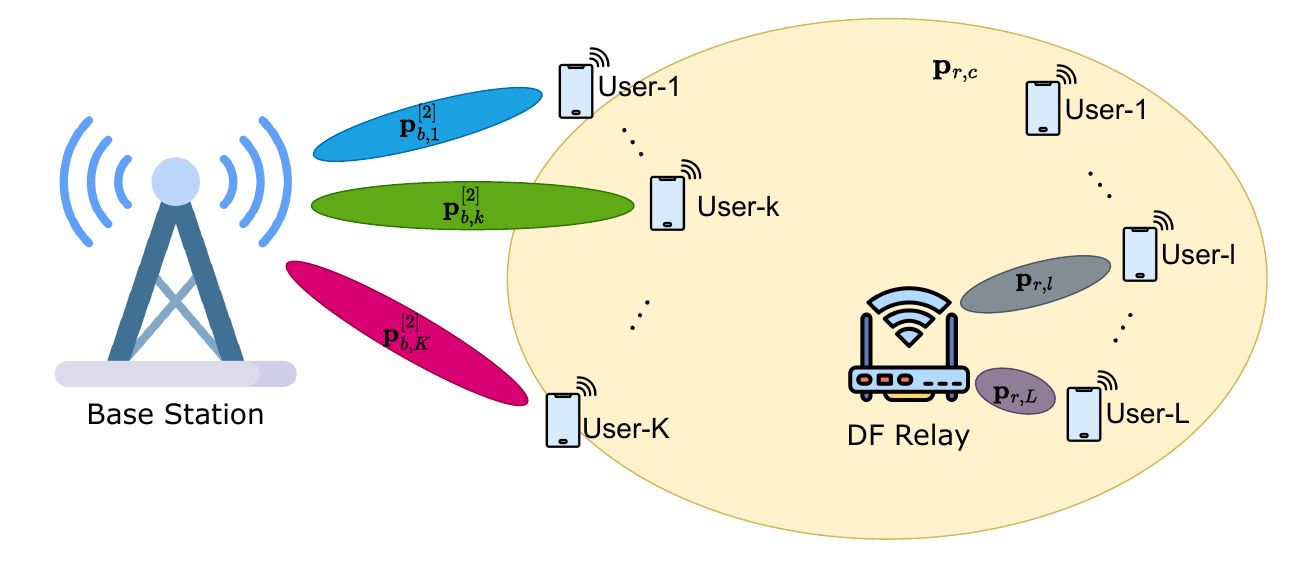}
            \caption[ Second phase.]%
            {{Second phase.}} 
            \label{subfig:secondphase}
        \end{subfigure}
        \hfill
        \caption[Transmission model.]
        {{\small Transmission model.}} 
\label{fig:Systemmodel}
    \end{figure*}
    
\section{System and Channel Models}
\label{section:System-model}
A two-phase multi-antenna communication system is considered where a \ac{DF} relay assists in serving multiple users using \ac{RSMA}. In the initial phase, as depicted in Fig. 1(\subref{subfig:firstphase}), the \ac{BS}, equipped with $N$ antennas, employs \ac{RSMA} to transmit both common and private streams to $K$ single-antenna \acp{BU} as well as the relay. During the subsequent phase, illustrated in Fig. 1(\subref{subfig:secondphase}), the relay utilizes its $M$ antennas to transmit both common and private streams to $L$ single-antenna \acp{RU} located within its coverage area but outside the \ac{BS} coverage. Simultaneously, the \ac{BS} enhances spectral efficiency by transmitting private streams to the \acp{BU}. Importantly, the \acp{BU} also receive the common stream broadcasted by the relay during this phase.

The relay utilizes digital precoding with $M$ dedicated \ac{RF} chains, each connected to an individual antenna. The \ac{BS} transmits $K$ streams to the \acp{BU} and $L$ streams to the \acp{RU}, where $L \leq M$ and $K+L \leq N$\footnote{For simplicity, we assume throughout this work that the system uses $L$ antennas for data transmission, where $L=M$. Through a straightforward generalization, the relay could also employ the remaining antennas (if $L<M$) to enhance spatial diversity.}. Notably, due to the inherent broadcasting nature of RSMA, the relay's common stream carries data intended for both \acp{RU} and \acp{BU}.

In the second phase, the \acp{BU} first decode the relay's common stream, then proceed to decode their private streams from the \ac{BS}. This approach enables the \ac{BS} to remain active and partially mitigate \ac{MUI} at the \acp{BU} through the relay's common stream, effectively forming a secondary \ac{RSMA} scheme. However, this scheme is inherently non-power optimal, as power optimization for common and private streams is not performed at a single source. These streams are transmitted from separate sources without coordinated optimization, and therefore, successful \ac{SIC} at the \acp{BU} during the second phase cannot be guaranteed.

In \ac{DF} relay systems, the performance is highly sensitive to the accuracy of \ac{CSIT}\footnote{In this paper, CSIT encompasses the channel knowledge at both the BS and the Relay.}, as precoder design depends on precise \ac{CSIT}. However, obtaining accurate \ac{CSIT} in practical scenarios is challenging due to factors such as channel estimation errors \cite{mao2018rate}, user mobility \cite{zhang2009mode}, and pilot contamination \cite{mishra2022mitigating}.
To facilitate the analysis, we adopt a generalized model that captures the relationship between the channel and the available \ac{CSIT}. In this model, the small-scale fading component of the \ac{BS} to \ac{BU} channel is represented as
\begin{equation}
    \label{eq:Imperfect_channel_of_users}
\begin{aligned}
    & \mathbf{h}_{b,k} = \sqrt{\epsilon^2} \mathbf{\widehat{h}}_{b,k} + \sqrt{1 - \epsilon^2} \mathbf{e}_{b,k},
\end{aligned}
\end{equation}
where $\mathbf{h}_{b,k}$ represents the spatially uncorrelated Rayleigh flat fading channel between the \ac{BS} and the $k$-th user, with i.i.d. entries following a complex Gaussian distribution $\mathcal{CN}(0,1)$, the terms $\mathbf{\widehat{h}}_{b,k}$ and $\mathbf{e}_{b,k}$ stand for the \ac{CSIT} and \ac{CSIT} error, respectively where both are are modeled as independent random variables with i.i.d. entries, also adhering to the $\mathcal{CN}(0,1)$ distribution. Similarly, the channels linking the relay to the $k$-th and $l$-th users are denoted as $\mathbf{h}_{r,k}$ and $\mathbf{h}_{r,l}$, respectively. 
The quality of the CSIT is quantified by the coefficient $\epsilon$, where $0 \leq \epsilon \leq 1$, and its value is influenced by practical system parameters such as the \ac{SNR}, user mobility speed, and the number of quantization bits employed for feedback.

For the channel between the \ac{BS} and the relay, we adopt a MIMO Rician fading model to incorporate the presence of a \ac{LOS} component, as described in \cite{benkhelifa2014low}. Accordingly, the Rician flat fading \ac{BS}-relay channel and corresponding \ac{CSIT} can be represented, respectively, as follows

\begin{equation}
    \begin{aligned}
        & \mathbf{H}_{b,r} =\sqrt{\frac{K_R}{K_R+1}} \overline{\mathbf{H}} + \sqrt{\frac{1}{K_R+1}} \mathbf{H}_{\omega}, 
    \end{aligned}
\end{equation}
\begin{equation}
    \begin{aligned}
        & \widehat{\mathbf{H}}_{b,r} = \sqrt{\frac{K_R}{K_R+1}} \overline{\mathbf{H}} + \sqrt{\frac{1}{K_R+1}} \widehat{\mathbf{H}}_{w},
    \end{aligned}
\end{equation}
where $K_R > 0$ represents the Rician factor, and $\overline{\mathbf{H}} \in \mathbb{C}^{N \times M}$ is a normalized deterministic complex matrix capturing the LOS component. The \ac{NLOS} component of the channel is modeled as $\mathbf{H}_{w} = \sqrt{\epsilon^2} \widehat{\mathbf{H}}_{w} + \sqrt{1 - \epsilon^2} \mathbf{E}_w$, where $ \widehat{\mathbf{H}}_{w}$ and $\mathbf{E}_w$ represent the estimated \ac{NLOS} component and its corresponding error, respectively. These matrices are assumed to be random and complex, with i.i.d. \ac{CSCG} entries, with zero mean and unit variance \cite{priya2023linear}. 

\section{Proposed Approach}
\label{sec:proposed}
\subsection{Proposed Transmission Model} 
In the first phase, the \ac{BS} employs \ac{RSMA} to transmit signals to $k$-th and $l$-th users via the relay. Following the \ac{RS} strategy, messages $W_{k}^{[1]}$\footnote{The superscript $(\cdot)^{[1]}$ and $(\cdot)^{[2]}$ are used to indicate elements associated with phase 1 and phase 2, respectively.} and $W_{l}$, intended for $k$-th and $l$-th users are split into common parts $W_{c,k}$ and $W_{c,l}$, and private parts $W_{p,k}^{[1]}$ and $W_{p,l}$, $\forall k \in K$ and $\forall l \in L$. The common parts are combined to form a common message $W_{c}^{[1]}$, which is subsequently encoded into the common stream $s_{c}^{[1]}$  using a codebook shared by users and the relay. The common stream $s_{c}^{[1]}$ contains partial information for each user. Hence, successful decoding of the common stream is essential for both \acp{BU} and \acp{RU} to retrieve their complete messages and the relay to facilitate transmission to the \acp{RU}\footnote{Perfect CSI at the receiver and \ac{SIC} are assumed throughout this work.}. Then, the private messages $W_{p,k}^{[1]}$ and $W_{p,l}$ are encoded into the private streams $s_{k}^{[1]}$ and $s_l^{[1]}$ using separate codebooks. Only $k$-th user can decode $s_k^{[1]}$, and the relay can decode $s_l^{[1]}$. The symbol vector encompassing the common and private streams is represented as $\mathbf{s} = [s_{c}^{[1]},..., s_{K}^{[1]},..., s_{L}^{[1]}]^T$, where $\mathbb{E}[\mathbf{s}\mathbf{s}^H] = \mathbf{I}$. The transmitted signal at the BS is shown as
\begin{equation}
\begin{aligned}
\label{eq:BS_TX_signal}
    \mathbf{x}_b^{[1]} =& \sqrt{P_1(1-t_1)}\mathbf{p}_{c,b} s_{c}^{[1]} \\ + &\sqrt{\frac{P_1 t_1}{K+L}} \left(\sum_{k \in K} \mathbf{p}_{b,k} s_{k}^{[1]} +  \sum_{l \in L} \mathbf{p}_{b,l} s_l^{[1]}\right), 
\end{aligned}
\end{equation}
where $\mathbf{p}_{c,b} \in \mathbb{C}^{N}$ represents the precoding vector for the common stream at the BS. Similarly, $\mathbf{p}_{b,k} \in \mathbb{C}^{N}$ and $\mathbf{P}_{b,r} = [\mathbf{p}_{b,1},...,\mathbf{p}_{b,l},...,\mathbf{p}_{b,L}] \in \mathbb{C}^{N \times L}$ denote the precoding vectors for the private streams intended for $k$-th user and the relay, respectively. We assume equal power allocation for the private streams\footnote{Equal power allocation among precoders is a widely adopted strategy in \ac{MU-MIMO} systems, both in academic research and practical implementations (see \cite{hao2015rate,dai2017multiuser,mishra2022mitigating} and related works).}, ensuring that $||\mathbf{p}_{c,b}||^2 = 1$,  $||\mathbf{p}_{b,k}||^2 = 1,  \forall k \in K $ and $||\mathbf{p}_{b,l}||^2 = 1,  \forall l \in L $. The power allocation coefficient $t_1$ governs the distribution of power between the common and the private streams, while $P_1$ represents the total transmit power at the \ac{BS}. When $t_1$ is set to 1, the common stream is deactivated, and the transmission strategy effectively reverts to \ac{SDMA} with equal power allocation among users. It is worth noting that since the relay applies precoding for the \acp{RU}, the \ac{BS} does not require \ac{CSIT} feedback from \acp{RU}. This reduces overhead at the \ac{BS}, leading to improved system efficiency.

The signals received at $k$-th user and the relay during the first phase are represented as $\mathbf{y}_k^{[1]} = \mathbf{h}_{b,k}^H \mathbf{x}_{b}^{[1]} + n_k$ and $\mathbf{y}_r = \mathbf{H}_{b,r}^H \mathbf{x}_{b}^{[1]} + \mathbf{n}_r$, respectively. 
Here, $n_k \sim \mathcal{CN}(0, \sigma_k^2)$ and $\mathbf{n}_r \in \mathbb{C}^{M } \sim \mathcal{CN}(0, \sigma_r^2\mathbf{I})$ represent the \ac{AWGN} at $k$-th user and the relay, respectively. Typical of an RSMA receiver, both $k$-th user and the relay decode the common stream first while treating all private streams as interference, then, employ \ac{SIC} to be able to decode the private stream. The remaining private streams are considered as noise while decoding the private stream of the $k$-th user \cite{yang2021optimization}.
Accordingly, the \ac{SINR} for the common and private streams at $k$-th user in the first phase are shown, respectively, as follows
\begin{equation}
\label{eq:rateKCphase1}
    \begin{aligned}
\gamma_{c,k}^{[1]} =&  \frac{ P_1\left(1-t_1\right) | \mathbf{h}_{b,k}^H \mathbf{p}_{c,b}|^2}{1 + \frac{P_1 t_1}{K+L}   \left( \sum_{j \in K}| \mathbf{h}_{b,k}^H \mathbf{p}_{b,j} |^2 + \sum_{l \in L}| \mathbf{h}_{b,k}^H \mathbf{p}_{b,l}|^2 \right)}, 
    \end{aligned}
\end{equation}
\begin{equation}
\label{eq:rateKPphase1}
    \begin{aligned}
\gamma_{k}^{[1]} =& \frac{ \frac{P_1 t_1}{K+L}  | \mathbf{h}_{b,k}^H \mathbf{p}_{b,k} |^2}{1 + \frac{P_1 t_1}{K+L}  \left( \sum\limits_{j \in K, j \neq k} |\mathbf{h}_{b,k}^H \mathbf{p}_{b,j} |^2 + \sum\limits_{l \in L}| \mathbf{h}_{b,k}^H \mathbf{p}_{b,l}|^2 \right)}.
    \end{aligned}
\end{equation}
The \acp{ER} of the common and private streams are indicated as $R_{c,k}^{[1]}(t_1) = \mathbb{E} \big\{\log_2 \big( 1+ \gamma_{c,k}^{[1]} \big)\big\}$ and $R_{k}^{[1]}(t_1) = \mathbb{E} \big\{\log_2 \big( 1+ \gamma_{k}^{[1]} \big)\big\}$, respectively, with the expectations taken over $\mathbf{h}_{b,k}$. \acp{ER} are utilized for system design in the presence of imperfect \ac{CSIT}, as the transmitter does not have access to instantaneous achievable rates under such conditions. This necessitates the reliance on \acp{ER} to account for channel uncertainty. Since the relay utilizes the \ac{DF} protocol to facilitate communication between the \ac{BS} and $l$-th user, the \ac{SINR} of both the common and private streams at the relay are expressed, respectively, as follows
\begin{equation}
\label{eq:rateRCphase1}
    \begin{aligned}
        \gamma_{c,l}^{[1]} =& \frac{ P_1(1-t_1)  |\mathbf{h}_{b,l}^H \mathbf{p}_{c,b}|^2}{1 + \frac{P_1 t_1}{K+L} \left( \sum_{k \in K} |\mathbf{h}_{b,l}^H \mathbf{p}_{b,k} |^2 +  \sum_{j \in L} | \mathbf{h}_{b,l}^H  \mathbf{p}_{b,j}|^2 \right)},
    \end{aligned}
\end{equation}
\begin{equation}
\label{eq:rateRPphase1}
    \begin{aligned}
        \gamma_{l}^{[1]} =& \frac{ \frac{P_1 t_1 }{K+L} |\mathbf{h}_{b,l}^H \mathbf{p}_{b,l} |^2}{1 + \frac{P_1 t_1}{K+L}  \left(\sum\limits_{k \in K} |\mathbf{h}_{b,l}^H \mathbf{p}_{b,k}|^2  + \sum\limits_{j \in L, j\neq l} |\mathbf{h}_{b,l}^H \mathbf{p}_{b,j} |^2\right)},
    \end{aligned}
\end{equation}
where $\mathbf{H}_{b,r} = [\mathbf{h}_{b,1},...,\mathbf{h}_{b,l},...,\mathbf{h}_{b,L}] \in \mathbb{C}^{N \times L}$, and the \acp{ER} of the common and private streams are represented as $R_{c,l}^{[1]}(t_1) = \mathbb{E} \big\{\log_2 \big( 1+ \gamma_{c,l}^{[1]}\big)\big\}$ and $R_{l}^{[1]}(t_1) = \mathbb{E} \big\{\log_2 \big( 1+ \gamma_{l}^{[1]} \big)\big\}$, respectively. 

In the first phase, the common stream rate allocated by considering  solely the \acp{BU}. This allocation choice is motivated by the expectation that the relay, equipped with multiple antennas and potentially benefiting from a stronger \ac{LOS} channel component, consistently possesses the capability to successfully decode the common stream. Consequently, we establish the common stream rate as $R_{c}^{[1]}(t_1) =  \min\{R_{c,k}^{[1]}(t_1)\}$. Our objective is to maximize the aggregate private stream rates encompassing both the \acp{BU} and the relay as well as common stream rate. We particularly emphasize maximizing the private stream rate at the relay, as it directly influences the rates achievable by the \acp{RU}, given that these streams are decoded and subsequently re-encoded at the relay before being forwarded. Therefore, the considered \ac{ESR} is used to determine the closed-form power allocation coefficient $t_1$ for the first phase is written as
\begin{equation}
\label{eq:ESR1}
    R^{[1]}(t_1) =  \frac{1}{2}\left(R_{c}^{[1]}(t_1) +  \sum_{k \in K} R_{k}^{[1]}(t_1) + \sum_{l \in L} R_{l}^{[1]}(t_1) \right),
\end{equation}
where the factor of $1/2$ accounts for the fact that symbols are transmitted for only half of the total time slot duration, as explained in \cite{liang2013limited}.

In the second phase, the relay re-encodes $s_{c}^{[1]}$ and $s_{l}^{[1]}$ into $s_{c}^{[2]}$ and $s_{l}^{[2]}$. Then, the relay transmits $s_{c}^{[2]}$ and $s_{l}^{[2]}$ to $l$-th user using the precoding vectors $\mathbf{p}_{c,r} \in \mathbb{C}^{L}$ and $\mathbf{p}_{r,l} \in \mathbb{C}^{L}$, ensuring that $||\mathbf{p}_{c,r}||^2 = ||\mathbf{p}_{r,l}||^2= 1$. Simultaneously, the BS transmits additional stream $s_{k}^{[2]}$ to \acp{BU}. 
Accordingly, the transmitted signals at the relay and BS during the second phase are shown, respectively, as
\begin{equation}
\begin{aligned}
    \mathbf{x}_r =& \sqrt{P_2(1-t_2)}\mathbf{p}_{c,r} s_{c}^{[2]} + \sqrt{\frac{P_2 t_2}{L}} \ \sum_{l \in L} \mathbf{p}_{r,l} s_{l}^{[2]}, 
\end{aligned}
\end{equation}
\begin{equation}
\begin{aligned}
        \mathbf{x}_b^{[2]} =&  \sqrt{\frac{P_1}{K}} \sum_{k \in K} \mathbf{p}_{b,k} s_{k}^{[2]}. 
\end{aligned}
\end{equation}
Then, the signal received at $l$-th user is represented as $\mathbf{y}_l = \mathbf{h}_{r,l} \mathbf{x}_{r} + n_l$ where $n_l \sim \mathcal{CN}(0, \sigma_l^2)$ represents the \ac{AWGN} affecting $l$-th user. Therefore, the \acp{ER} for the common and private streams at $l$-th user are shown, respectively, as follows
\begin{equation}
\label{eq:rateLCphase2}
    \begin{aligned}
R_{c,l}^{[2]}(t_2) =&  \mathbb{E} \left\{\log_2 \left( 1+ \frac{ P_2(1-t_2) | \mathbf{h}_{r,l}^H \mathbf{p}_{c,r} |^2}{1 + \frac{P_2 t_2}{L}  \sum_{j \in L} |\mathbf{h}_{r,l}^H \mathbf{p}_j |^2} \right)\right\},  
    \end{aligned}
\end{equation}
\begin{equation}
\label{eq:rateLPphase2}
    \begin{aligned}
R_{l}^{[2]}(t_2) =& \mathbb{E} \left\{\log_2 \left( 1+ \frac{ \frac{P_2 t_2 }{L}  | \mathbf{h}_{r,l}^H \mathbf{p}_l|^2}{1 + \frac{P_2 t_2}{L}  \sum_{j \in L j \neq l} |\mathbf{h}_{r,l}^H \mathbf{p}_j |^2} \right)\right\}.
    \end{aligned}
\end{equation}
We prioritize the \acp{RU} when determining the common stream rate, as the primary responsibility of relay is to serve those users. Thus, the common stream rate in the second phase is set to $R_{c}^{[2]}(t_2) = \min\{R_{c,l}^{[2]}(t_2)\}$. Due to this architecture, there may be residual interference if \acp{BU} are unable to completely decode the common stream. This residual interference can be modelled and accounted into the computation of the private stream rates for \acp{BU} during the second phase of transmission. Such a circumstance can be described as imperfect \ac{SIC} resulting from imperfect CSI at the receiver. 
However, this stems directly from omitting \acp{BU}' rate requirements in the relay's common stream rate allocation. Therefore, the received signal of the $k$-th user in the second phase is shown as 
\begin{equation}
    \begin{aligned}
        \mathbf{y}_k^{[2]} & = \mathbf{h}_{b,k}  \mathbf{x}_b^{[2]} +  \mathbf{h}_{r,k} \mathbf{x}_r + n_k, \\
        & =   \sqrt{P_2(1-t_2)} \mathbf{h}_{r,k} 
        \mathbf{p}_{c,r} (\hat{s}_{c,k} + s_{c,k,res})  \\
        & + \! \sqrt{\frac{P_1}{K}} \sum_{k \in K} \mathbf{h}_{b,k} \mathbf{p}_{b,k} s_{k}^{[2]} \! +  \! \sqrt{\frac{P_2 t_2}{L}} \ \sum_{l \in L} \mathbf{h}_{r,k} \mathbf{p}_{r,l} s_{l}^{[2]} \!+\! n_k, \\
    \end{aligned}
\end{equation} 
where $\hat{s}_{c,k}$ and $s_{c,k,res}$ denote achievable and residual common stream rate at the $k$-th user, respectively. The received signal of the $k$-th user after applying \ac{SIC} is modelled as
\begin{equation}
    \begin{aligned}
        & \mathbf{y}_{k,\text{SIC}}^{[2]} =   \mathbf{y}_k^{[2]} - \mathbf{h}_{r,k} \mathbf{p}_{c,r} \hat{s}_{c,k} = \sqrt{P_2(1-t_2)} \mathbf{h}_{r,k} 
        \mathbf{p}_{c,r} s_{c,k,res} \\
        & +   \sqrt{\frac{P_1}{K}} \sum_{k \in K} \mathbf{h}_{b,k} \mathbf{p}_{b,k} s_{k}^{[2]} + \sqrt{\frac{P_2 t_2}{L}} \ \sum_{l \in L} \mathbf{h}_{r,k} \mathbf{p}_{r,l} s_{l}^{[2]} + n_k. \\
    \end{aligned}
\end{equation} 
Based on this, the \ac{SINR} for the common and private streams at $k$-th user in the second phase are shown, respectively, as 
\begin{equation}
    \begin{aligned}
    \label{eq:Rate_2_KC}
         \gamma_{c,k}^{[2]} =&  \frac{ P_2(1-t_2) |\mathbf{h}_{r,k}^H \mathbf{p}_{c,r} |^2}{1 + \frac{P_1}{K}  \sum_{j \in K} |\mathbf{h}_{b,k}^H \mathbf{p}_j |^2 + \frac{P_2 t_2}{L} \sum_{l \in L}  | \mathbf{h}_{r,k}^H  \mathbf{p}_l|^2}, 
    \end{aligned}
\end{equation} 
\begin{equation}
    \begin{aligned}
    \label{eq:Rate_2_KP}
         \gamma_{k}^{[2]} = \frac{ \frac{P_1}{K} | \mathbf{h}_{b,k}^H \mathbf{p}_k|^2}{ 1+ \frac{P_1}{K}  \sum\limits_{j \in K j \neq k} |\mathbf{h}_{b,k}^H \mathbf{p}_j |^2  + \frac{P_2 t_2}{L} \sum\limits_{l \in L}  | \mathbf{h}_{r,k}^H  \mathbf{p}_l|^2 + I_{k,res}}. 
    \end{aligned}
\end{equation} 
Therefore, the \acp{ER} of the common and private streams are indicated as $R_{c,k}^{[2]}(t_2) = \mathbb{E} \big\{\log_2 \big( 1+ \gamma_{c,k}^{[2]} \big)\big\}$ and $R_{k}^{[2]}(t_2) = \mathbb{E} \big\{\log_2 \big( 1+ \gamma_{k}^{[2]} \big)\big\}$, respectively, where $I_{k,res}$ is the residual common stream interference, which is derived as
\begin{subequations}
\label{eq:I_R derivation}
\begin{align}
I_{k,res} 
 = &\mathbb{E}\{ |\sqrt{P_2(1-t_2)}  \mathbf{h}_{r,k} \mathbf{p}_{c,r} s_{c,k,res}|^2\} \label{subeq:GMI},\\
= & P_2(1-t_2)  \mathbb{E}\{ 
|\mathbf{h}_{r,k}|^2\} \mathbb{E}\{|\mathbf{p}_{c,r}|^2
\} \mathbb{E}\{|s_{c,k,res}|^2\} \label{subeq:Independence},\\
= & P_2(1-t_2) \sigma_{s_{c,k,res}}^2 \label{subeq:Unit variance} = P_2(1-t_2) \gamma_{c,k,res},\\
= & P_2(1-t_2) (2^{R_{c,k,res}}-1) \label{subeq:Rate},\\
= & P_2(1-t_2) (2^{R_{c}^{[2]}-\min\{R_{c,k}^{[2]},R_{c}^{[2]}\}}-1) \label{subeq:Rate Difference},
\end{align}
\end{subequations}
where \eqref{subeq:Independence} is derived from the \ac{GMI} analysis as presented in \cite{lee2022max}. The independence in \eqref{subeq:Independence} occurs because $\mathbf{p}_{c,r}$ is built for $\mathbf{h}_{r,l}$, and hence independent of $\mathbf{h}_{r,k}$. In \eqref{subeq:Unit variance}, we assume unit power for both the common precoder and the channel. Then, the residual \ac{SNR} is expressed in terms of rate in \eqref{subeq:Rate}, following the approach in \cite{kotaba2020urllc}. In \eqref{subeq:Rate Difference}, the residual rate is formulated based on rate differences, and $\min\{R_{c,k}^{[2]}(t_2),R_{c}^{[2]}(t_2)\}$ is taken, as the common stream rate achievable at the \acp{BU} is limited by the relay's common stream rate. In the second phase, the total rate for \acp{RU} is maximized under relay priority. Then, the \ac{ESR} is used to determine the closed-form power allocation coefficient $t_2$ for the second phase as follows
\begin{equation}
\label{eq:ESR2}
    R^{[2]}(t_2) = \frac{1}{2} \left(R_{c}^{[2]}(t_2) +  \sum_{l \in L} R_{l}^{[2]}(t_2)\right).
\end{equation}

\subsection{Total Sum-Rate Expressions}

Inherent to \ac{DF} relay systems, the achievable rate at the relay inherently constrains the maximum attainable sum-rate of the \acp{RU}, as established in \cite{host2005capacity}. Therefore, the achievable \ac{ESR} for the users is expressed as 

\begin{equation}
    \begin{aligned}
    \label{eq:rate_realistic}
         R_{1}(t_1,t_2) \! = \! \frac{1}{2}\bigg(R_{c}(t_1,t_2)\!+\!\sum_{l\in L} R_{l}(t_1,t_2)\!+\!\sum_{k \in K} R_k(t_1,t_2) \bigg),
    \end{aligned}
\end{equation} 
where 

\begin{subequations}
\begin{align}
    & R_k(t_1,t_2) = R_{k}^{[1]}(t_1) + R_{k}^{[2]}(t_2) \label{eq:Total_Rate_BS_Users}, \\ 
    &   R_{c}(t_1,t_2) = \min\{ R_{c}^{[1]}(t_1), R_{c}^{[2]}(t_2)\} \label{eq:Common_rate_limit},\\ 
    &  R_{l} \left(t_1,t_2\right) = \min\{ R_{l}^{[1]}(t_1),  R_{l}^{[2]}(t_2)\}, \forall \ l \in L, 
    \label{eq:Private_rate_limit}
\end{align}
\end{subequations}
where \eqref{eq:Total_Rate_BS_Users} represents the aggregate private stream rates of \acp{BU} across both phases, while \eqref{eq:Common_rate_limit} and \eqref{eq:Private_rate_limit} denote the common and private stream rate limitations imposed by the relay, respectively\footnote{While the second-phase common stream contributes to enhanced system reliability through diversity gain, this gain is not explicitly factored into the sum-rate expressions for analytical tractability.}.
Due to the inherent variability of Rayleigh fading channels experienced by both the RUs and BUs, constraints on the achievable common stream rate, $R_{c}(t_1,t_2)$, can originate from either $R_{c}^{[1]}(t_1)$ or $R_{c}^{[2]}(t_2)$. Since $R_{c}^{[1]}(t_1)$ is allocated to cater to all users and hence cannot be modified, we focus on addressing potential limitations arising from $R_{c}^{[2]}(t_2)$. One strategy is reallocating the second-phase common stream at the relay by removing \acp{BU} data and reallocating that rate to the \acp{RU}, potentially ensuring successful decoding for \acp{BU} in the first phase and \acp{RU} in the second phase. Therefore, the common stream rate contributing to the overall sum-rate originates solely from the first phase, leading to the following ESR formulation
\begin{equation}
    \begin{aligned}
    \label{eq:rate_realistic_BS_C}
         R_{2} (t_1,t_2)\! = \! \frac{1}{2} \bigg(R_{c}^{[1]}(t_1)\! +\! \sum_{l \in L} R_{l} \left(t_1,t_2\right)\! +  \!\sum_{k \in K} R_k(t_1,t_2) \bigg).
    \end{aligned}
\end{equation}

In a case where the private stream rate limitation of the relay is ignored and considering solely the common stream rate limitation, the theoretical \ac{ESR} $R_{3}(t_1,t_2)$ is provided as
\begin{equation}
    \begin{aligned}
    \label{eq:common_rate_limitations}
         R_{3} (t_1,t_2) \!= \! \frac{1}{2} \bigg(R_{c}(t_1,t_2)\! + \!\sum_{l \in L} R_{l}^{[2]}(t_2) \!+ \! \sum_{k \in K} R_k(t_1,t_2)\bigg).
    \end{aligned}
\end{equation} 
Similarly, in the case of the absence of relay
rate limitations, the theoretical ESR $R_{4} \left(t_1,t_2\right)$ is defined as
\begin{equation}
    \begin{aligned}
    \label{eq:without_limitaiton}
         R_{4} (t_1,t_2)\! =\!  \frac{1}{2} \bigg(R_{c}^{[1]}(t_1) \!+\! \sum_{l \in L} R_{l}^{[2]}(t_2)\! +\! \sum_{k \in K} R_k(t_1,t_2)\bigg).
    \end{aligned}
\end{equation}

\section{Lower Bound Analysis}
\label{sec:lowerbound}
Our goal is to find power allocation strategies that maximize the \ac{ESR} of \ac{RSMA} in \eqref{eq:ESR1} and \eqref{eq:ESR2}, considering imperfect CSIT. We provide a feasible lower bound for the 1-layer RSMA sum rate in two phases, to achieve a closed-form solution for optimal power coefficients $t_{1,opt}$ and $t_{2,opt}$. Henceforth, these approximated ESRs will be referred to as AESRs.

\subsection{Lower Bound for First Phase}
The AESR of the first phase can be given as
\begin{equation}
\label{eq:lower_bound_first_phase}
    \Tilde{R}^{[1]}(t_1) =  \frac{1}{2}\left(\Tilde{R}_{c}^{[1]}(t_1) +  \sum_{k \in K} \Tilde{R}_{k}^{[1]}(t_1) + \sum_{l \in L} \Tilde{R}_{l}^{[1]}(t_1) \right),
\end{equation}
where $\Tilde{R}_{c}^{[1]}(t_1)$, $\Tilde{R}_{k}^{[1]}(t_1)$ and $\Tilde{R}_{l}^{[1]}(t_1)$ represent the \acp{AER}, serving as approximations for $R_{c}^{[1]}(t_1)$, $R_{k}^{[1]}(t_1)$ and $R_{l}^{[1]}(t_1)$, respectively, as defined in \eqref{eq:ESR1}. We conduct the derivation in two parts: first, a lower bound for the sum of the \acp{AER} of the private streams in the first phase is derived, represented as $\sum_g^G \Tilde{R}_g(t_1) = \sum_{k \in K} \Tilde{R}_{k}^{[1]}(t_1) + \sum_{l \in L} \Tilde{R}_{l}^{[1]}$, second, a lower bound for the \ac{AER} of the common stream is derived, denoted as $\Tilde{R}_c(t_1)$.

\subsubsection{Lower Bound for $\sum_g^G \Tilde{R}_g(t_1)$}
The \acp{SINR} of private streams as shown in \eqref{eq:rateKPphase1} and \eqref{eq:rateRPphase1}, are represented comprehensively within the \ac{ER} expression as

\begin{equation}
\label{eq:rateGphase1}
    \begin{aligned}
         R_{g}^{[1]}(t_1) = \mathbb{E} \left\{ \log_2 \left( 1+ \frac{ \frac{P_1 t_1}{G} | \mathbf{h}_{b,g}^H \mathbf{p}_{g} |^2}{1 + \frac{P_1 t_1}{G}  \sum_{j \in G j \neq g} |\mathbf{h}_{b,g}^H \mathbf{p}_j |^2} \right)\right\}, 
    \end{aligned}
\end{equation}
where $G = K+L$, $\mathbf{h}_{b,g}$ represents the channel between the BS and $k$-th user, $\forall g \in K$, and the channel between the BS and $l$-th user, $\forall g \in L$. Likewise, $\mathbf{p}_{g}$ denotes the ZF precoding vector for $k$-th user, $\forall g \in K$, and for $l$-th user, $\forall g \in L$.
 
To start, we provide several helpful features that are essential to determining the boundaries. Consider a \ac{ZF} precoding is employed among the \acp{BU} and the relay, the precoder $\mathbf{p}_{g}$ for $k$-th user, is selected to be orthogonal to the channel of $j$-th user, $\forall k, j \in K$, $k \neq j$, and to the channel of $l$-th user, $\forall l \in L$. Similarly, for $l$-th user, the precoder is orthogonal to the channel of $i$-th user, $\forall l,i \in L, i \neq l$, and to the channel of $k$-th user, $\forall k \in K$.
 
Due to the isotropic nature of i.i.d. Rayleigh fading, \ac{DoF} are reduced to $N - G + 1$ at the \ac{BS}. Therefore, for \ac{ZF} precoders with equal power allocation, the term $|\sqrt{2}\widehat{\mathbf{h}}_g^H \mathbf{p}_{g}|^2$ follows a chi-squared distribution with \ac{DoF} $2(N-G+1)$, i.e., $|\sqrt{2}\widehat{\mathbf{h}}_g^H \mathbf{p}_{g}|^2 \sim \chi^2_{2(N - G + 1)}$. Equivalently, $|\widehat{\mathbf{h}}_g^H \mathbf{p}_{g}|^2 \sim \text{Gamma}(N-G+1,1)$ \cite{jindal2010multi}. Additionally, the isotropic distribution of the ZF precoder $\mathbf{p}_{g}$ and its independence from the Gaussian error $\mathbf{e}_j^H$, $\forall j,g \in G$, result in $|\mathbf{e}_k^H \mathbf{p}_{g}|^2 \sim \text{Gamma}(1,1)$ \cite{underloaded}. Finally, a second-order moment-matching method is utilized to approximate the sum of Gamma-distributed random variables. Given $Y = \sum_i A_i$, where each $A_i \sim \text{Gamma}(D_i, \theta_i)$ and $\widehat{D} = \frac{(\sum_i D_i \theta_i)^2}{\sum_i D_i \theta_i^2}$, $\widehat{\theta} = \frac{\sum_i D_i \theta_i^2}{\sum_i D_i \theta_i}$ \cite{jaramillo2014coordinated}. Initially, a lower bound for the \ac{ER} of an arbitrary receiver-$g$, denoted as $R_g^{[1]}(t_1)$ is derived. Subsequently, we generalize this lower bound to encompass all $G$ receivers, thereby obtaining the lower bound for $\sum_g^G \Tilde{R}_g(t_1)$ \cite{underloaded}. The updated Gamma parameters for the first phase, obtained through the moment-matching method, are expressed as
\begin{equation}
    \begin{aligned}
    \label{eq:Gamma_parameters_G}
    \widehat{D}_g =& \frac{[\epsilon^2(N+1) + (1-2\epsilon^2)G] ^2}{\epsilon^4(N+1)+(1-2\epsilon^2)G}, 
    \\
    \widehat{\theta}_g =& \frac{\epsilon^4(N+1)+(1-2\epsilon^2)G}{\epsilon^2(N+1) + (1-2\epsilon^2)G}, 
        \end{aligned}
\end{equation}
and the generalized the lower bound of private streams given in \eqref{eq:lower_bound_first_phase} to $G$ receivers with independent channels having identical second order statistics as

\begin{equation}
\begin{aligned}
\label{eq:Bound_Rate_Phase_1_Private}
        \sum_{g=1}^{G} \tilde{R}_g(t_1) & \geq G\log_2 \left(1 + \frac{P_1   e^{\mu_g}}{G} t_1 \right) \\
      - & G\log_2 \left(1 + \frac{P_1  (1-\epsilon^2)(G-1)}{G} t_1\right),
\end{aligned}
\end{equation}
where $\mu_g = \ln(\widehat{\theta}_g) + \Gamma'(\widehat{D}_g)/\Gamma(\widehat{D}_g)$. 
The proofs of \eqref{eq:Gamma_parameters_G} and \eqref{eq:Bound_Rate_Phase_1_Private} are derived from \cite[Lemma 1]{underloaded} and \cite[Proposition 1]{underloaded}, respectively. The detailed proof is omitted for brevity.

\subsubsection{Lower Bound for $\Tilde{R}^{[1]}_c(t_1)$}
\label{subsubsec:R_c_1}
 We define r.v. $Y_k$ from \eqref{eq:rateKCphase1} as
\begin{equation}
    Y_k = \frac{| \mathbf{h}_{b,k}^H \mathbf{p}_{c,b}|^2}{1 + \frac{P_1 t_1}{G}   \left( \sum_{j \in K}| \mathbf{h}_{b,k}^H \mathbf{p}_{b,j} |^2 + \sum_{l \in L}| \mathbf{h}_{b,k}^H \mathbf{p}_{b,l}|^2 \right)},
\end{equation}
where $Y^{[1]} = \min_{k \in K} Y_k$. For tractability of the expressions, we approximate the r.v. $Y^{[1]}$ by a r.v. $\Tilde{Y}^{[1]}$. The approximated common stream rate in the first phase can be expressed as follows
\begin{subequations}
\begin{align}
     \Tilde{R}_c(t_1) & \geq \mathbb{E} \left\{ \log_2(1+P_1(1-t_1)\Tilde{Y}^{[1]}) \right\} \\
    &  \geq  \log_2(1+P_1(1-t_1)\mathbb{E} \left\{ \Tilde{Y}^{[1]}) \right\} \label{eq:Jensen}\\
    &  \geq  \log_2\left(1+P_1(1-t_1) e^{\mathbb{E} \left\{ \ln{(\Tilde{Y}^{[1]})} \right\}} \right)  \label{eq:Concave}.
\end{align}
\end{subequations}
The expressions \eqref{eq:Jensen} and \eqref{eq:Concave} are derived from Jensen's inequality, utilizing the properties that $\log_2(1+ax)$ is a concave function of $x$ for any $ax > 0$ and $\log_2(1 + ae^x)$ is a convex function of $x$ for any $a>0$, respectively \cite{hao2015rate}.
In order to obtain $\mathbb{E} \{\ln{(\Tilde{Y}^{[1]})}\}$, the value of $\beta_g = \mathbb{E}\{\ln{(\Tilde{Y}^{[1]})}\}$ in \eqref{eq:Concave} is derived from \cite[Lemma 3]{underloaded}, yielding the following result
\begin{equation}
    \begin{aligned}
    \label{eq:beta_1}
     & \beta_g =  - \gamma - \ln(K) - e^{\frac{KG}{P_1 \widehat{\theta}_g t_1}} \sum_{m=1}^{\nint{\widehat{D}_gK}} \mathrm{E}_{m}\left(\frac{KG}{P_1 \widehat{\theta}_g t_1}\right). \\
    \end{aligned}
\end{equation}
By combining the lower bounds for $\sum_g^G \Tilde{R}_{g}(t_1)$ and $\Tilde{R}_c(t_1)$, the AESR of the first phase can be lower bounded as
\begin{equation}
\label{eq:Bound_Rate_Phase_1}
\begin{aligned}
        &\Tilde{R}^{[1]}(t_1) \geq - G\log_2 \left(1 + \frac{P_1  (1-\epsilon^2)(G-1)}{G} t_1\right)  \\
     & + G\log_2 \left(1 + \frac{P_1   e^{\mu_g}}{G} t_1 \right) + \log_2 \left(1 + P_1  (1-t_1)  e^{\beta_g}\right) . 
\end{aligned}
\end{equation}

\subsection{Lower Bound for Second Phase}
The AESR of the second phase can be given as
\begin{equation}
\label{eq:lower_bound_second_phase}
    \Tilde{R}^{[2]}(t_2)  =  \frac{1}{2}\left(\Tilde{R}_{c}^{[2]}(t_2) +  \sum_{l \in L} \Tilde{R}_{l}(t_2) \right),
\end{equation}
where $\Tilde{R}_{c}^{[2]}(t_2)$ and $\Tilde{R}_{l}(t_2)$ represent the \acp{AER} of $R_{c}^{[2]}(t_2)$ and $R_{l}^{[2]}(t_2)$, respectively, as defined in \eqref{eq:ESR2}.
 
\subsubsection{Lower Bound for $\sum_l^L \Tilde{R}_{l}(t_2)$}
To establish a lower bound for the \ac{AER} of $l$-th user private stream rate, denoted as $\Tilde{R}_{l}(t_2)$, we leverage the fact that $|\widehat{\mathbf{h}}_l^H \mathbf{p}_{l}|^2 \sim \text{Gamma}(M - L + 1)$ \cite{jaramillo2014coordinated}. Furthermore, due to the isotropic ZF precoder $\mathbf{p}_{l}$ being independent of the Gaussian error $\mathbf{e}_j^H$, we have $|\mathbf{e}_l^H \mathbf{p}_{l}|^2 \sim \text{Gamma}(1,1)$ and $\sum_{j \in L,  j \neq l}  |\mathbf{h}_{r,l}^H \mathbf{p}_j |^2  \sim \text{Gamma}(L-1,1)$. Then, the updated Gamma parameters at the second phase, obtained using the moment-matching method, are given by
 
\begin{equation}
    \begin{aligned}
    \label{eq:Gamma_parameters_L}
    \widehat{D}_l =& \frac{[\epsilon^2(M+1) + (1-2\epsilon^2)L] ^2}{\epsilon^4(M+1)+(1-2\epsilon^2)L}, 
    \\ 
    \widehat{\theta}_l =& \frac{\epsilon^4(M+1)+(1-2\epsilon^2)L}{\epsilon^2(M+1) + (1-2\epsilon^2)L}.
    \end{aligned}
\end{equation}

We can generalize the lower bound of private streams given in \eqref{eq:lower_bound_second_phase} to $L$ users with independent channels having identical second order statistics as

\begin{equation}
\begin{aligned}
\label{eq:sum-rate-of-second-phase_private}
        \sum_{l=1}^{L} \tilde{R}_{l}(t_2) & \geq L\log_2 \left(1 + \frac{P_2   e^{\mu_l}}{L} t_2 \right) \\
      - & L\log_2 \left(1 + \frac{P_2  (1-\epsilon^2)(L-1)}{L} t_2\right),
\end{aligned}
\end{equation}
where $ \mu_l = \ln(\widehat{\theta}_l) + \Gamma'(\widehat{D}_l)/\Gamma(\widehat{D}_l)$. The proofs of \eqref{eq:Gamma_parameters_L} and \eqref{eq:sum-rate-of-second-phase_private} are derived from \cite[Lemma 1]{underloaded} and \cite[Proposition 1]{underloaded}, respectively. The detailed proof is omitted for brevity.

\subsubsection{Lower Bound for $\Tilde{R}_c^{[2]} (t_2)$}
We define r.v. $Y_l$ from \eqref{eq:rateLCphase2} as
\begin{equation}
    Y_l = \frac{|\mathbf{h}_{r,l}^H \mathbf{p}_{c,r} |^2}{1 + \frac{P_2 t_2}{L}  \sum_{j \in L} |\mathbf{h}_{r,l}^H \mathbf{p}_j |^2},
\end{equation}
where $Y^{[2]} = \min_{l \in L} Y_l$. To tractability of the expressions, we approximate the r.v. $Y^{[2]}$ by a r.v. $\Tilde{Y}^{[2]}$. Following a similar approach as in \ref{subsubsec:R_c_1}, a lower bound for the \ac{AER} $\Tilde{R}_c^{[2]} (t_2)$ is derived as
\begin{equation}
\begin{aligned}
     \Tilde{R}_c^{[2]} (t_2)  \geq  \log_2\left(1+P_2(1-t_2) e^{\mathbb{E} \left\{ \ln{(\Tilde{Y}^{[2]})} \right\}} \right), 
\end{aligned}
\end{equation}
where $\beta_l = \mathbb{E}\{\ln{(\Tilde{Y}^{[2]})}\}$ and it is defined as
\begin{equation}
    \begin{aligned}
      & \beta_l = - \gamma - \ln\left(L\right) - e^{\frac{L^2}{P_2 \widehat{\theta}_l t_2}} \sum_{m=1}^{ \widehat{D}_l L} \mathrm{E}_{m}\left(\frac{L^2}{P_2 \widehat{\theta}_l t_2}\right). \\
    \end{aligned}
\end{equation}
By combining the lower bounds for $\sum_l^L \Tilde{R}_{l}(t_2)$ and $\Tilde{R}_c^{[2]}(t_2)$, the AESR of the second phase can be lower bounded as
\begin{equation}
\label{eq:Bound_Rate_Phase_2}
\begin{aligned}
        &\Tilde{R}^{[2]}(t_2) \geq  - L\log_2 \left(1 + \frac{P_2  (1-\epsilon^2)(L-1)}{L} t_2\right) \\
        & + \log_2 \left(1 + P_2  (1-t_2)  e^{\beta_l}\right)  
       + L\log_2 \left(1 + \frac{P_2   e^{\mu_l}}{L} t_2 \right).   
\end{aligned}
\end{equation}

\section{Closed-Form Solution for the Optimal Power Allocation}
\label{section:Closed Form}
This section focuses on deriving the optimal power allocation coefficients, $t_{1}$ and $t_2$, that maximize the lower bounds in \eqref{eq:Bound_Rate_Phase_1} and \eqref{eq:Bound_Rate_Phase_2}. Given the computational complexity of an exhaustive search, a closed-form solution for the optimal coefficients, $t_{1,opt}$ and $t_{2,opt}$, is developed for practical implementation.
 
\subsection{Derivation for First Phase}
Leveraging the assumption that $P_1t_1 \rightarrow \infty$, the final term in \eqref{eq:beta_1} is reformulated as follows, with a similar derivation presented in \cite[Proposition 4]{underloaded}
 
\begin{equation}
\begin{aligned}
        & e^{\frac{KG}{P_1\widehat{\theta}_gt_1}} \sum_{m=1}^{\nint{\widehat{D}_gK}} E_m \left( \frac{KG}{P_1 \widehat{\theta}_g t_1} \right) \\
        &\approx \ln \left( \nint{\widehat{D}_gK}-1 \right) + \frac{1}{2 \left( \nint{\widehat{D}_gK}-1 \right)} -\ln\left( \frac{KG}{P_1 \widehat{\theta}_g t_1} \right).
\end{aligned}
\end{equation}
Then, $\beta_g$ is approximated as

\begin{equation}
    \begin{aligned}
        \beta_g \approx - \gamma   + \ln\left( \frac{G}{P_1 \widehat{\theta}_g (\nint{\widehat{D}_gK}-1) t_1} \right) - \frac{1}{2 \left( \nint{\widehat{D}_gK}-1 \right)}.
    \end{aligned}
\end{equation}
Based on the assumption $P_1t_1 \rightarrow \infty$ the lower bound expression \eqref{eq:Bound_Rate_Phase_1} can be written as
\begin{equation}
\begin{aligned}
\label{eq:to_derivative_1}
         &\Tilde{R}^{[1]}(t_1) \geq  G\log_2 \left(1 + \frac{P_1   e^{\mu_g}}{G} t_1 \right) \\
      & - G\log_2 \left(1 + \frac{(1-\epsilon^2)(G-1)}{G} t_1\right) \\
      & + \log_2 \left( 1 + \frac{(1-t_1) }{t_1}\frac{G}{ \widehat{\theta}_g (\nint{\widehat{D}_g K} - 1) } e^{-\gamma - \frac{1}{2(\nint{ \widehat{D}_g K} - 1)}} \right) \\
        = & - G \log_2 \left( \frac{1}{t_1 \tau_g} + \frac{\omega_g}{\tau_g}\right) - \log_2 \left(1 - \rho_g + \frac{\rho_g}{t_1} \right),
\end{aligned}
\end{equation}
with the terms 
\begin{equation}
    \begin{aligned}
    & \tau_g = \frac{P_1  e^{\mu_g} }{G}, \ 
    \omega_g = \frac{P_1 (1-\epsilon^2)(G-1)}{G}
    \\
    & \rho_g = \frac{G}{\widehat{\theta}_g(\nint{\widehat{D}_gK}-1)}  e^{- \gamma  - \frac{1}{2 \left( \nint{\widehat{D}_g K}-1 \right)}}.
    \end{aligned}
\end{equation}
By differentiating \eqref{eq:to_derivative_1} with respect to $t_1$ and setting the resulting expression equal to zero, $t_{1,opt}$ is obtained as

\begin{equation}
    \begin{aligned}
        \label{eq:t_opt_1}
        t_{1,opt} = \frac{\rho_g (G-1)}{ \rho_g(\omega_g+G)-G}.
    \end{aligned}
\end{equation}

To ensure a valid value for $t_{1,opt}$, specifically within the range $(0, 1]$, the condition $\rho_g(\omega_g+G)-G > \rho_g (G-1)$ is imposed. Then, the power allocation algorithm in the first phase is expressed as
    
\begin{equation}
    \begin{aligned}
        t_{1,opt} = \begin{cases}
        \frac{\rho_g (G-1)}{ \rho_g(\omega_g+G)-G}, & \text{if } \rho_g (\omega_g + 1) / G > 1 \\
        1, & \text{otherwise}.
    \end{cases}
\end{aligned}
\end{equation}

\subsection{Derivation for Second Phase}
Employing a similar methodology, the lower bound expression \eqref{eq:Bound_Rate_Phase_2} is rewritten as

\begin{equation}
    \begin{aligned}
    \label{eq:to_derivative_2}
    &\Tilde{R}^{[2]}(t_2) \geq  - L \log_2 \left( \frac{1}{t_2 \tau_l } + \frac{\omega_l}{\tau_l}\right) - \log_2 \left(1 - \rho_l + \frac{\rho_l}{t_2} \right), \\
    \end{aligned}
\end{equation}
with the terms 
\begin{equation}
    \begin{aligned}
    & \tau_l = \frac{P_2 e^{\mu_l} }{L}, \ 
    \omega_l = \frac{P_2 (1-\epsilon^2)(L-1)}{L}
    \\
    & \rho_l = \frac{L}{\widehat{\theta}_l(\nint{\widehat{D}_lL}-1)}  e^{- \gamma  - \frac{1}{2 \left( \nint{\widehat{D}_lL}-1 \right)}}. 
    \end{aligned}
\end{equation}
By differentiating \eqref{eq:to_derivative_2} with respect to $t_2$ and setting the resulting expression equal to zero, $t_{2,opt}$ is obtained as

\begin{equation}
    \begin{aligned}
        t_{2,opt} = \frac{\rho_l (L-1)}{ \rho_l(\omega_l+L)-L}.
    \end{aligned}
\end{equation}

To ensure a valid value for $t_{2,opt}$, specifically within the range $(0, 1]$, the condition $\rho_l(\omega_l+L)-L > \rho_l (L-1)$ is imposed. Then, the power allocation algorithm in the second phase is expressed as

\begin{equation}
    \begin{aligned}
    \label{eq:t_opt_2}
    t_{2,opt} = \begin{cases}
        \frac{\rho_l (L-1)}{ \rho_l(\omega_l+L)-L}, & \text{if } \rho_l (\omega_l + 1) / L > 1 \\
        1, & \text{otherwise}.
    \end{cases}
\end{aligned}
\end{equation}

\section{Simulation Results}    
\label{section:Simulation Results}
This section presents numerical results demonstrating the efficacy of the proposed scheme, with performance evaluated through Monte Carlo simulations over 200 channel realizations. Two types of CSIT error are considered; scaling error, where error variance $\epsilon^2$ decreases with increasing SNR as $\epsilon^2 = 1-P_i^{-\tau}, i \in 1,2$ \cite{mao2018rate,mishra2021rate}, and non-scaling error, where $\epsilon^2$ remains constant. The scaling error model is relevant when CSIT quality improves with higher SNR, typically due to channel estimation errors. The parameter $\tau$ controls this improvement rate. Conversely, the constant error model captures the impact of factors like mobility, latency, and pilot contamination that hinder CSIT quality improvement even with increased SNR \cite{zhang2009mode,truong2013effects,mishra2022mitigating}. The default antenna and user configuration is $N=16$, $M=8$, $K=8$, $L=8$, and $K_R$ set to 10, unless otherwise specified. To maximize common stream performance for both phase, we utilize the leftmost eigenvector (associated with the maximum eigenvalue) of the corresponding CSIT matrix for common stream precoding.

Fig. \ref{fig:Closed-Form vs Exhaustive Search} illustrates the achievable ESRs of the proposed RSMA scheme under various relay rate limitations and channel imperfection scenarios. Different cases are considered as follows:

\begin{figure}
        \centering
        \begin{subfigure}[b]{0.475\textwidth}
            \centering
\includegraphics[width=\textwidth]{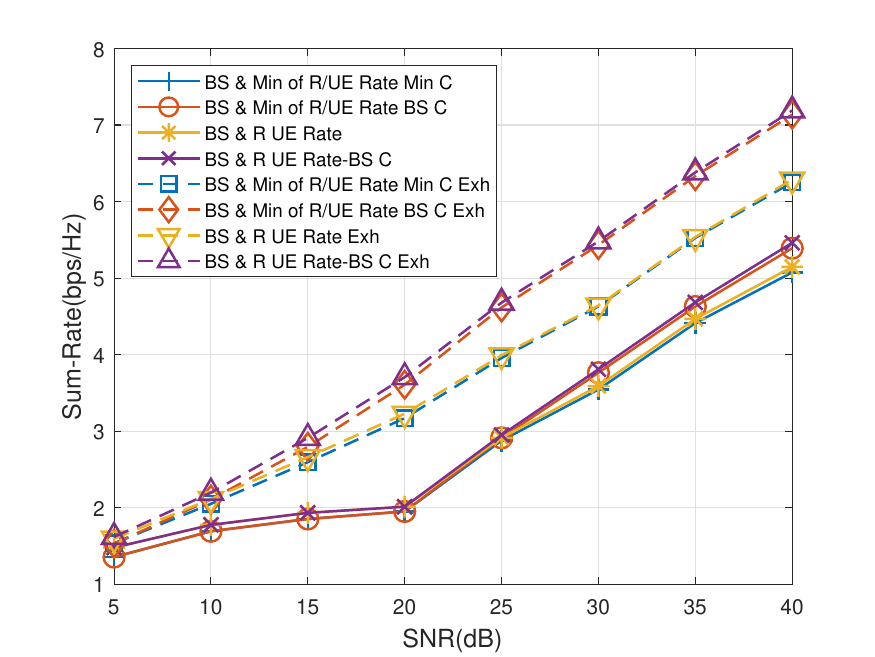}
            \caption[Constant error = 0.7]%
            {{\small Constant error ($\epsilon=0.3$) }} 
            \label{subfig:Constant_07_1}
        \end{subfigure}
        \hfill
        \begin{subfigure}[b]{0.475\textwidth}   
            \centering \includegraphics[width=\textwidth]{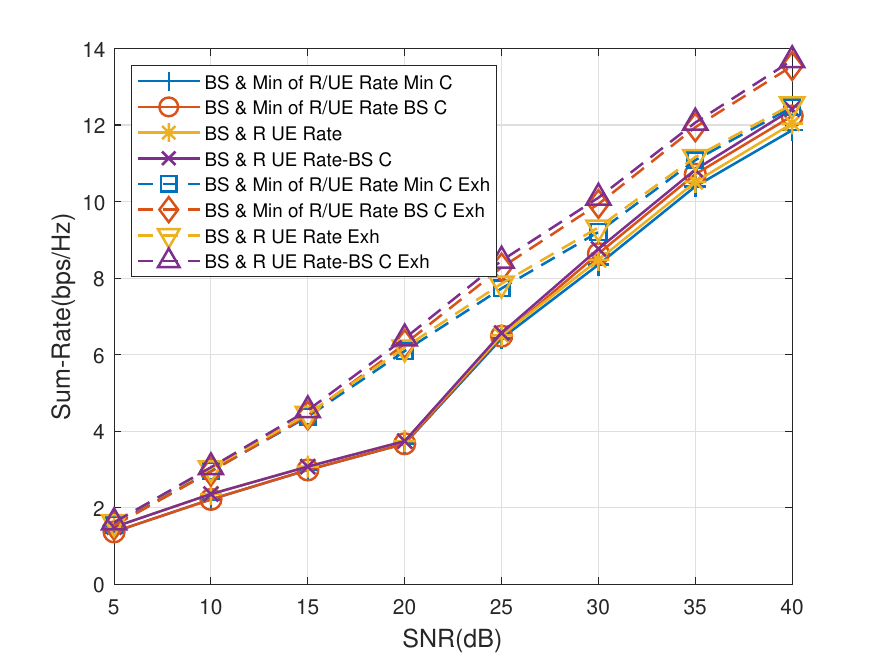}
            \caption[ Scaling error ($\epsilon=\sqrt{1-P^{-0.1}}$)]%
            {{\small Scaling error ($\epsilon=\sqrt{1-P^{-0.1}}$)}} 
            \label{subfig:Scaling_05_1}
        \end{subfigure}
        \hfill
        \caption[ Closed-form vs exhaustive search ]
        {{\small Closed-form RSMA vs exhaustive search with different relay transmission limitations.}} 
\label{fig:Closed-Form vs Exhaustive Search}
    \end{figure}

\begin{itemize}
    \item BS \& Min of R/UE Rate: Sum-rate considering both relay common and private streams rate limitations from both phases given in \eqref{eq:rate_realistic}.
    \item BS \& Min of R/UE Rate BS C: Sum-rate accounting for RU private stream rate limitations and common stream reallocation for RUs given in \eqref{eq:rate_realistic_BS_C}.
    \item BS \& R UE Rate: Sum-rate with only common stream rate limitation given in \eqref{eq:common_rate_limitations}.
    \item BS \& R UE Rate-BS C: Idealized sum-rate without any limitations given in \eqref{eq:without_limitaiton}.
\end{itemize}

\begin{figure}
        \centering
        \begin{subfigure}[b]{0.475\textwidth}
            \centering
\includegraphics[width=\textwidth]{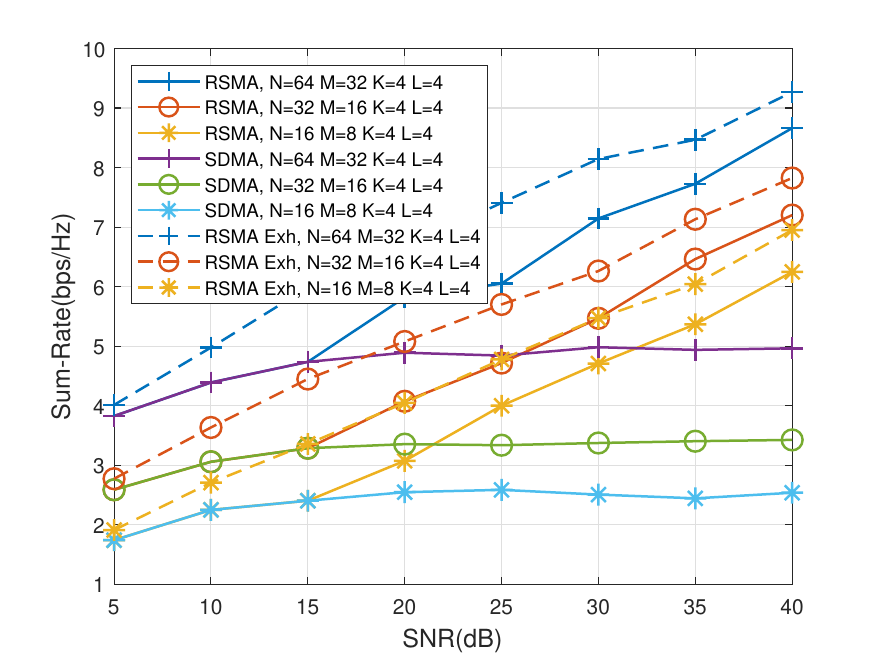}
            \caption[Constant error = 0.3]%
            {{\small Constant error ($\epsilon=0.3$) }} 
            \label{subfig:Antenna_Constant_07}
        \end{subfigure}
        \hfill
        \begin{subfigure}[b]{0.475\textwidth}   
            \centering \includegraphics[width=\textwidth]{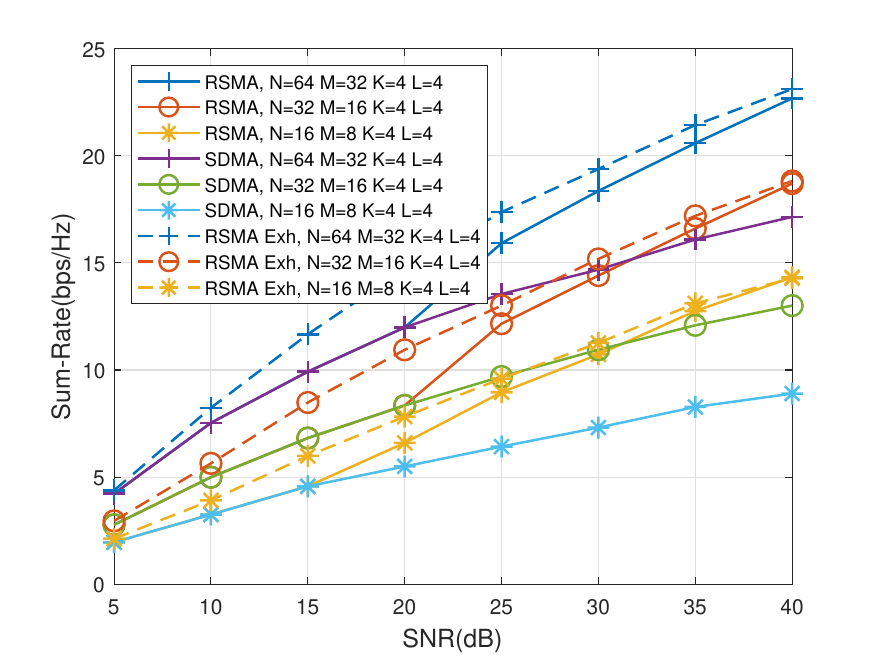}
            \caption[ Scaling error ($\epsilon=\sqrt{1-P^{-0.1}}$)]%
            {{\small Scaling error ($\epsilon=\sqrt{1-P^{-0.1}}$)}} 
            \label{subfig:Antenna_Scaling_01}
        \end{subfigure}
        \hfill
        \caption[ RSMA vs SDMA with different antenna configurations. ]
        {{\small RSMA vs SDMA with different antenna configurations.}} 
\label{fig:Different Antenna Configurations}
    \end{figure}
    
For comparison, we include results from an exhaustive search over $t_1,t_2 \in [10^{-6},1]$ with adaptive resolutions in Figs. 2(\subref{subfig:Constant_07_1})-2(\subref{subfig:Scaling_05_1}). While our design prioritizes BUs in the first phase and excludes them from second-phase power allocation, leading to a slightly sub-optimal overall performance, both the exhaustive search and proposed closed-form methods achieve comparable sum-rates, particularly for scaling errors as shown in Fig. 2(\subref{subfig:Scaling_05_1}). This highlights the near-optimality and low complexity of our closed-form solution. In contrast, the exhaustive search evaluates all possible $t_1$ and $t_2$ combinations at a given resolution to find the sum-rate maximizing one. Slightly lower rates are achievable at the relay with the given Rician factor, assuming sorted and matched relay and \acp{RU} private stream rates. The marginal gain from removing the relay rate constraint suggests its achievable rate usually surpasses that of its users. However, due to channel randomness, either BUs or RUs can become the bottleneck for the common stream rate. Therefore, choosing the BS common stream rate using the scheme in \eqref{eq:rate_realistic_BS_C}, which accounts for practical channel constraints, achieves the maximum sum-rate, albeit with the additional complexity of reallocating the common stream rate at the relay.

  \begin{figure}
        \centering
        \begin{subfigure}[b]{0.475\textwidth}
            \centering
\includegraphics[width=\textwidth]{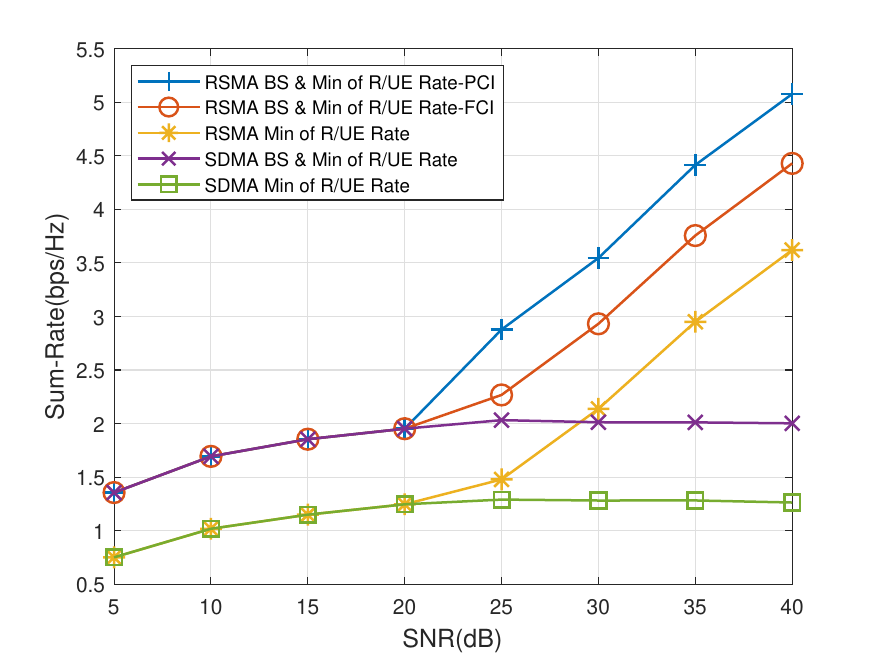}
            \caption[Constant error = 0.3]%
            {{\small Constant error ($\epsilon=0.3$)}} 
            \label{subfig2:Constant_07_2}
        \end{subfigure}
        \hfill
        \begin{subfigure}[b]{0.475\textwidth}   
            \centering \includegraphics[width=\textwidth]{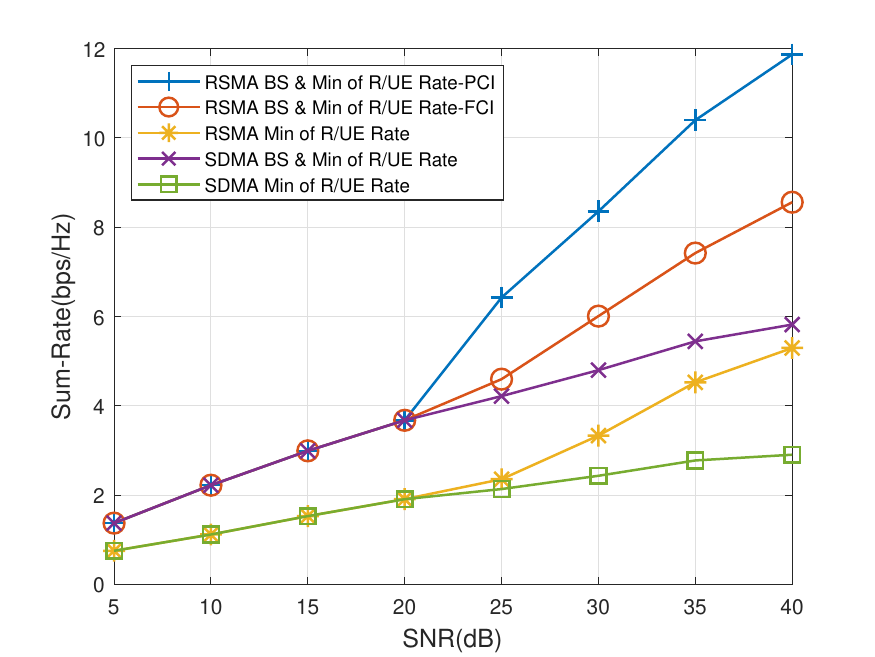}
            \caption[ Scaling error = -0.1]%
            {{\small Scaling error ($\epsilon=\sqrt{1-P^{-0.1}}$)}} 
            \label{subfig:Scaling_05_2}
        \end{subfigure}
        \caption[ RSMA vs SDMA Both Phase ]
        {{\small Impact of the proposed RSMA scheme at the \acp{BU} during the second phase on the total achievable ESR.}} 
\label{fig:RSMA vs SDMA Both Phase}
    \end{figure}

Furthermore, we benchmark RSMA against SDMA, employing both the proposed and exhaustive search power allocation methods, to highlight superior performance of RSMA in DF-assisted MU-MIMO RSMA systems. Figs. 3(\subref{subfig2:Constant_07_2})-3(\subref{subfig:Scaling_05_2}) compares the performance of the proposed method and SDMA with constant and scaling error, respectively, with varying $N, M, K, L,$ and $\epsilon^2$, covering both small-scale and massive MIMO scenarios. Sum-rate of SDMA drops significantly with increasing channel error and saturates at high SNRs, indicating an interference-limited regime. In contrast, RSMA achieves a non-saturating rate in all configurations. The gain from RSMA depends on the number of transmit antennas at BS and at relay, and CSIT quality $\epsilon^2$. Higher $N$ and $M$ lead to reduced gains as CSIT quality improves. However, improving CSIT quality is challenging in practice for large $N$ and $M$ due to CSI feedback overhead and pilot contamination. Thus, RSMA offers a significant performance advantage for any combination of $N, M, K, L,$ and $\epsilon^2$ in practical MU-MIMO systems. Therefore, the proposed RSMA scheme exhibits robust performance in massive MIMO scenarios, consistently achieving sum-rates comparable to exhaustive search across different antenna configurations.

\begin{figure}[t]
    \centering
    \includegraphics[width=1.0\linewidth]{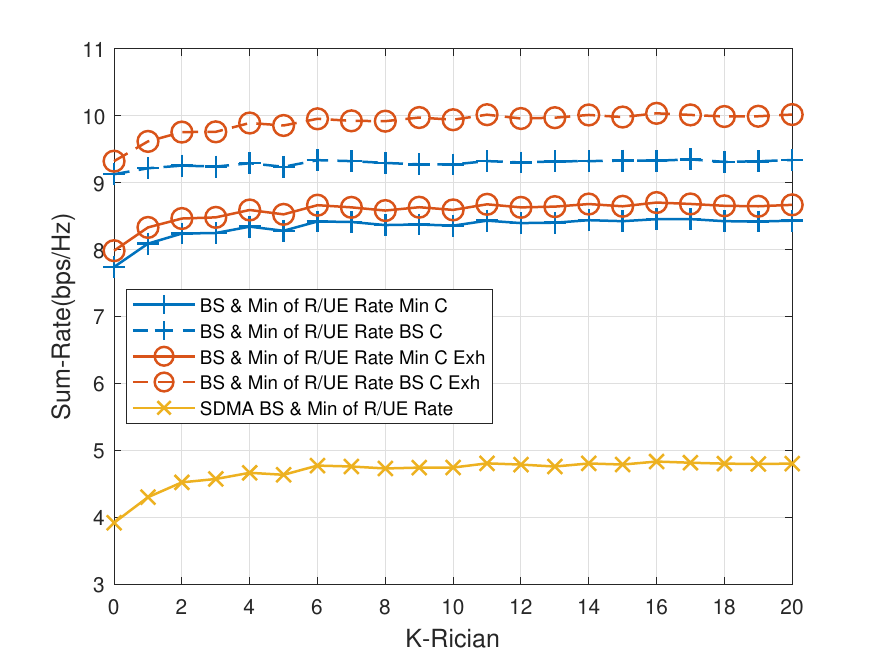}
    \caption{Impact of K-Rician fading on the overall sum-Rate.}
    \label{fig:K_Rician}
\end{figure}

Fig. \ref{fig:RSMA vs SDMA Both Phase} compares the impact of partially decoding (\ac{PCI}) the relay's common stream versus treating it as pure interference (\ac{FCI}), or not considering  the BUs in the second phase at all. Fig. 4(\subref{subfig2:Constant_07_2}) shows a notable difference of approximately 15\% (0.64 bps/Hz) at high SNR regime when decoding the common stream compared to treating it fully as interference, and a 40\% difference (1.44 bps/Hz) when partially decoding the common stream at the BUs and not considering BUs in the second phase at all\footnote{To ensure a fair comparison, any remaining antennas at the relay are utilized for diversity.}. Both \ac{FCI} and \ac{PCI} approaches perform similarly until 20 dB, as our power allocation method does not allocate power to the common part, effectively reducing RSMA to SDMA. Fig. 4(\subref{subfig:Scaling_05_2}) shows a more substantial difference of about 38\% (3.3 bps/Hz) when decoding the common stream compared to fully treating it as interference, and a 119\% difference (6.62 bps/Hz) when partially decoding the common stream at the BUs and not considering BUs in the second phase at all. This highlights the importance of not underestimating the achievable common stream rate at BUs, even when the common stream is not explicitly allocated for them in the second phase, as it can significantly improve the total sum-rate.

Fig. \ref{fig:K_Rician} presents a comparative analysis of the sum-rates achieved by various transmission strategies under different K-Rician factors for the BS-relay channel, thereby elucidating the influence of channel quality on the performance of a two-hop relay system. In this analysis, a scaling error model with $\epsilon=\sqrt{1-P^{-0.1}}$ is employed to capture the impact of imperfect CSIT. It is observed that as the K-Rician factor increases, signifying a stronger \ac{LOS} component and improved channel quality, the sum-rates for both relay transmission strategies converge beyond a K-Rician factor of 5. This convergence suggests that the relay's capacity ceases to be a limiting factor for the overall sum-rate, even when the relay's achievable rate is not explicitly considered in the first phase common stream rate allocation. Notably, the proposed RSMA scheme consistently exhibits superior performance compared to SDMA across the entire range of K-Rician factors. Furthermore, the proposed scheme demonstrates near-optimal performance when benchmarked against the exhaustive search method, particularly for the sum-rate expression given in \eqref{eq:rate_realistic}.

\section{Conclusion and Future Work}
\label{section:Conclusion}
In conclusion, this paper conducted a comprehensive investigation into the implementation of RSMA within the context of multi-user DF relay systems under imperfect CSIT. A key contribution lies in deriving an analytical expression to quantify residual interference at BUs when decoding the common stream of relay. Additionally, we proposed a relay strategy where the common stream rate is actively reallocated to address imperfect CSIT challenges. We formulated a sum-rate optimization problem for the two-phase DF relay RSMA system and derived tractable lower bounds for the AESRs, leading to closed-form power allocation algorithms. Extensive simulations validated effectiveness of our approach in mitigating the impact of imperfect CSIT, highlighting potential of RSMA as a scalable and resilient solution for future relay-aided wireless systems.  

As a future work, we intend to enhance the proposed approach by explicitly incorporating diversity gains into the system design and analysis. Specifically, we plan to investigate the integration of RS in DF relay systems with link-level simulations to evaluate the performance gains achievable through various diversity combining schemes in the presence of imperfect CSIT.

\section*{Acknowledgment}
The authors would also like to thank Onur Dizdar for
his continuous support and guidance.


\vfill
\end{document}